# In search of a theory of supercooled liquids

## STEVEN A. KIVELSON[1] AND GILLES TARJUS[2]

[1]Department of Physics, Stanford University, Stanford, California 94305-2035, USA
[2]Laboratoire de Physique Théorique de la Matière Condensée, Universite P. and M. Curie, 4 Place Jussieu, 75252 Paris Cedex 05, France
e-mail: kivelson@stanford.edu; tarjus@lptl.jussieu.fr

Despite the absence of consensus on a theory of the transition from supercooled liquids to glasses, the experimental observations suggest that a detail-independent theory should exist.

It is widely accepted that the glass transition is one of the deepest and most important unsolved problems in condensed matter physics. It is also one of the oldest. The glass transition is one of the most spectacular phenomena in all of physics in terms of dynamical range: as the temperature is reduced over the modest interval from the melting temperature, $T_m$, to the glass transition, $T_g$, the shear-viscosity and the structural relaxation time can increase by 15 orders of magnitude. (To put this in perspective, the temperature change $(T_m - T_g)/T_m$ is typically around 30%.) It should be cause for general embarrassment in the field that there is still no consensus on even the most basic aspects of the theory of this phenomenon — whether the physics can be understood as arising from the modest growth of as yet undetected thermodynamic correlations that are then enormously amplified in terms of their consequences for the dynamics, or whether instead it is a purely dynamical phenomenon with little or no relation to any feature of the thermodynamics.

Nevertheless, there are good reasons to believe that there exists a detail-independent collective description of the salient properties of so-called 'fragile' supercooled liquids as they approach the glass transition. Assuming that such a theory exists, we explore the constraints on its character implied by existing experimental observations.

## IS A GENERAL THEORY POSSIBLE?

Many phenomena in nature are specific, and depend on details. In discussing 'the theory' of supercooled liquids, we are implicitly assuming that there is something about the phenomena that is independent of the microscopic details. Renormalization group theory accounts for the 'universality' of phenomena as the critical point at which a phase transition occurs is approached. There, the existence of a large correlation length implies that all microscopic differences are averaged away. If a weaker version of the same sort of detail-independence pertains to the glass transition, it probably is also a consequence of an emergent length scale that is long compared with the size of a molecule.

> It should be cause for general embarrassment in the field that there is still no consensus on even the most basic aspects of the theory of this phenomenon

We will discuss the evidence of a supermolecular length scale, $L$, in glass-forming liquids, but, even optimistically interpreted, this length is never found to be more than 5 or 10 molecular diameters. It is plausible that this is long enough that a collective description, based on the assumed proximity of the system to a critical point of one sort or another, may be reasonable. We will accept this assumption as a working hypothesis. Certainly, as we enumerate various phenomena below, the reader cannot fail to be struck by the apparent similarities among observations in quite a range of different liquids. However, because $L$ is not enormous, one must expect that, at best, a general theory can only be semiquantitatively compared with experiment.

In all supercooled liquids and polymers[1–4], the viscosity, $\eta$, increases exponentially with decreasing temperature, as

$$\eta(T) \approx \eta_\infty \exp[\Delta(T)/T], \qquad (1)$$

where $\eta_\infty$ is the viscosity in the limit of high temperature, and $\Delta(T)$ is an effective activation energy. The slowest local relaxation processes are characterized by a time, $\tau_\alpha$, that is proportional to $\eta$. In 'strong' glass-formers[1] such as $SiO_2$, where $\Delta(T)$ is nearly temperature-independent between $T_m$ and $T_g$, this behaviour has no obvious collective or 'cooperative' origin. However, other glass-formers[1] are defined to be 'fragile' if $\Delta(T)$ increases significantly with decreasing temperature, as shown in Fig. 1. This 'super-Arrhenius' temperature dependence is suggestive of a growing correlation length.

In the following, we will summarize some additional evidence that the dynamical arrest in fragile glass-formers is collective and shows more or less universal features. Two characteristics of the data[1–4] are particularly suggestive of this conclusion. First, there are common features to the temperature dependence of $\Delta(T)$ that permit various forms of scaling analysis. Second, the time dependence of the relaxation functions is not simply exponential, but again is sufficiently similar in different liquids to permit approximate data collapse. These observations make a plausible case that a 'general theory' is possible.

## THE IMPORTANT TEMPERATURE

There is no consensus concerning what specific temperature characterizes the important collective phenomena. The temperature $T_g$ is the extrinsically determined temperature at which the time to reach local equilibration exceeds our patience. It is the most important temperature from a practical standpoint, as it separates the glass from the liquid. But it is clearly irrelevant from the standpoint of the fundamental physics, because its value depends on the rate at which the liquid is cooled. (Because of the extraordinarily strong $T$ dependence of $\tau_\alpha$, in practice the rate dependence of $T_g$ is weak.) The melting temperature $T_m$ is also irrelevant; it is the essence of good glass-formers that, when supercooled, they do not explore the regions of configuration space corresponding to the crystalline order.

Most theories invoke an important characteristic temperature (see Fig. 2). Many envisage a true, but in practice unattainable, phase transition would occur at a temperature $T_0 < T_g$, if the experiments were carried out sufficiently slowly that local equilibrium could be maintained[5–8]. Presumably, this dynamically unattainable transition would be a thermodynamic transition from a supercooled liquid to a state referred to as an 'ideal glass'. It has also been suggested[9–11] that there is a well-defined crossover temperature, $T^*$, at which the characteristic collective behaviour evinced by the supercooled liquid begins. This crossover could be thermodynamic[10], associated with a narrowly avoided phase transition ($T^* \geq T_m$), or it could be a purely dynamical onset[11] of collective congestion. There is a class of 'mode-coupling' theories that envisage a crossover temperature, $T_c$, between $T_m$ and $T_g$ at which the dominant form of the dynamics changes[12]. Finally, there are models and theories in which the only characteristic temperature scale is microscopic, but there is a zero-temperature dynamical[11,13] or thermodynamical[14] critical point, which, although experimentally unattainable, is responsible for the interesting physics.

## IMPORTANT THERMODYNAMICAL FACTS

For those theories that envisage a fundamentally thermodynamic origin of the collective congestion in supercooled liquids, the most discouraging fact is that there is no clear evidence of any growing thermodynamic correlation length. On the other hand, existing experiments only measure the density–density (pair) correlation function, so if the putative order is of a more subtle type, perhaps it could have eluded detection. Attempts to measure multipoint correlations are obviously of central importance, but they have not been successful so far.

Conversely, there are two observations that are challenging for those theories with no fundamental involvement of thermodynamics. The first is the famous Kauzmann paradox[15]. The excess entropy, $\Delta S$, which is defined as the difference between the entropies of the supercooled liquid and the crystal, is a strongly decreasing function of $T$ from $T_m$ to $T_g$ and extrapolates to 0 at a temperature, $T_K$, which, for fragile glass-formers, is only 20–30% below $T_g$. Even though the crystal is, as we argued above, not relevant to the physics of the supercooled liquid, there is a sensible rationale for considering $\Delta S$. Most fragile glass-formers are molecular liquids in which a significant fraction of the entropy is associated with intramolecular motions. By subtracting the entropy of the crystal, one hopes to eliminate most of the contributions from extraneous degrees of freedom. A large change in the entropy is something to be taken very seriously.

The second observation is that there is an empirical relation between $\Delta S$ and the slow dynamics[5,16]. Specifically, there seems to be a correlation between the decrease of $\Delta S(T)$ and the increase of $\Delta(T)$ with decreasing temperature.

## IMPORTANT DYNAMICAL FACTS

The most important experimental fact about fragile, supercooled liquids is the super-Arrhenius growth of $\eta$ and $\tau_\alpha$ (see Fig. 2). Several kinds of functional fits to the $T$ dependence of $\eta$ and $\tau_\alpha$ have been presented, each motivated by a different theoretical prejudice concerning the underlying physics.

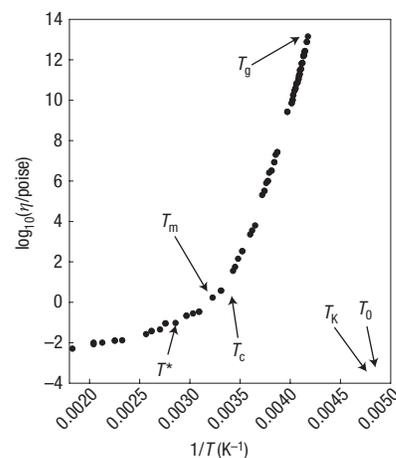

**Figure 2** Temperature-dependent viscosity of *ortho*-terphenyl, with the various possibly important temperatures indicated by arrows. (Several approaches take $T = 0$ as the only relevant temperature.) The original data are taken from references listed in refs 3 and 4.

A popular fit to the data over a range of temperature from somewhat below $T_m$ down to $T_g$ is achieved with the Vogel–Fulcher–Tammann (VFT) form, $\Delta(T) = DT[T_0/(T − T_0)]$, where $D$ is a fitting parameter, with its implication of the existence of an 'ideal glass transition' at $T_0 < T_g$ where $\eta$ and $\tau_\alpha$ would diverge. In a somewhat narrower range of temperatures, but with fewer adjustable parameters, a comparably good fit to the data is obtained with a power-law formula[17] $\Delta(T) = E_0[E_0/T]$, which diverges only at $T = 0$. A somewhat better global fit over the whole available range of temperature, but with one more free parameter than the VFT equation, is achieved with a form suggested by 'avoided critical behaviour' around a crossover temperature $T^*$ (ref. 10). Certainly, none of the above formulae fit the data perfectly, but all fit it as well as could be expected, so it does not seem possible to establish the validity of one over the other on the basis of the relatively small deviations between the fits and experiment.

It is also important to realize that the growth of the effective activation barrier $\Delta(T)$ is neither a divergent effect, nor a small one (Fig. 1); in some fragile liquids (for example, *ortho*-terphenyl), $\Delta(T_g)$ is roughly 3 or 4 times its high-temperature value, $\Delta_\infty$. This observation suggests that, whatever activation barrier is being surmounted in the key relaxation processes, near $T_g$ many molecules must move cooperatively.

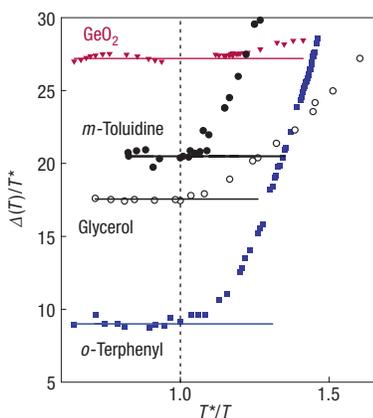

**Figure 1** Temperature-dependent effective activation energy of several supercooled liquids (see equation (1)) in units of the empirically determined crossover temperature scale, $T^*$. *Ortho*-terphenyl is one of the most 'fragile' glass-forming liquids, whereas GeO₂ is relatively strong. The original data are taken from references listed in refs 3 and 4.

In supercooled liquids, for $T < T_m$, the relaxation functions are distinctly non-exponential, and the relaxation spectra are non-Lorentzian[1–4]. In particular, at long times the relaxation functions can be well approximated by a 'stretched exponential', $\Phi(t) \approx \exp[-(t/\tau_a)^\beta]$. In the normal liquid regime, $T > T_m$, the stretching parameter $\beta$ is independent of $T$ and is close to 1 (although in most molecular glass-formers $\beta$ is actually closer to 0.8). For $T_m > T > T_g$, $\beta$ decreases gently down to $\beta \approx 0.3–0.5$ at $T = T_g$ (refs 1–4). The deviation from exponential relaxation (that is, the magnitude of $1 - \beta$) seems to correlate with the fragility (that is, the extent of the deviation from Arrhenius behaviour in the $T$ dependence of $\eta$ and $\tau_a$; refs1–4,4,18), although there are insufficient systematic data to quantify this correlation.

## SUPERMOLECULAR LENGTH SCALES

In the past decade, experimental[19,20,23] and numerical[21,22] evidence has accumulated of the existence of a supermolecular length scale in supercooled liquids and polymers associated with 'dynamic heterogeneities'. Specifically, the notion is that over moderately long times (although still not long compared with relaxation times), spatially localized regions of the liquid relax much more quickly than the average. (Of course, on long enough timescales, no location behaves differently than any other.)

Several important questions arise in this context: Is there a unique supermolecular length scale in the problem? Is the length scale purely dynamic, or is there a corresponding thermodynamic correlation length? Does this length, in fact, have anything to do with the super-Arrhenius slowing down and the broad distribution of relaxation rates?

Because most of the estimates yield only modest length scales (5–10 molecular diameters at $T_g$), these questions may be difficult to answer, even in principle. Because the relevant thermodynamic correlations, if they exist, are likely to be subtle[8,10], there is essentially no existing experiment of a thermodynamic quantity that is expected to show the same supermolecular scale.

Finally, although there is little systematic information on the dependence of the length scale on fragility, dynamical heterogeneities seem to have been observed in strong as well as fragile glass-formers[19,20,23,24]. On the face of it, this would seem to imply that the mere existence of a supermolecular dynamical length scale does not, by itself, lead to the special properties of fragile liquids on which we have focused.

It is clear that further investigations of the nature of the heterogeneities in supercooled liquids represent the forefront of the experimental effort in this field. It would be helpful to have more systematic comparative studies of strong and fragile glass-formers, to explore the temperature dependence of these phenomena over as broad a range of temperature as possible, and to find ways of exploring thermodynamic correlations of similar character to the interesting dynamic ones.

## CONCLUDING REMARKS

To the extent that the theory of supercooled liquids is, indeed, detail-independent, it should apply to many diverse systems. Thus, it is in principle useful to broaden the scope of the systems under study. Proposed analogies have been drawn with jammed systems including foams and granular materials, and with electronic systems with competing interactions. However, we have rather specific phenomena in mind, and before getting carried away with broad-based comparative studies, it is important to ascertain whether the analogy is deep or superficial.

It is a glaring omission in the present analysis that we have considered only the properties of the liquid at $T > T_g$, and have ignored the glass itself. There are many well-studied and general properties of glasses that include two-level systems, the 'boson peak', nonlinear relaxation and ageing. Because the glass is, more or less, a frozen liquid, it is clear that these properties reflect, in some way, the structure of the parent supercooled liquid. However, most of these properties do not depend in any obvious way on whether the liquid was fragile or strong, so it is not clear what constraints can be placed on a collective theory of fragile supercooled liquids based on experiments in the glass.

The set of essential experimental facts that we have discussed are probably not sufficient to resolve the deep issues of perspective that exist in the field. What we feel is needed is a strategy for exaggerating the relevant collective properties, so that asymptotically precise statements can be verified or falsified. In terms of collective theories, what is needed is an analytically or numerically tractable model system in which the limit of 'extreme fragility' could be realized. Specifically, one would like to access a regime in which any presumed supermolecular length scale, be it dynamical or thermodynamical, is arbitrarily large compared with the molecular scale. Of course, finding a real liquid that is much more fragile than the prototypical example, *ortho*-terphenyl, which appears in our figures, would be the greatest boon of all.